%% file: paper.tex
\documentclass[sigconf,nonacm]{acmart}

\usepackage{csquotes}
\usepackage{pdflscape}
\usepackage{color}
\usepackage{listings}
\usepackage{booktabs}
\usepackage{wasysym}
\usepackage{xspace}
\usepackage{titlesec}

\definecolor{codegreen}{rgb}{0,0.6,0}
\definecolor{codegray}{rgb}{0.5,0.5,0.5}
\definecolor{codepurple}{rgb}{0.58,0,0.82}
\definecolor{backcolour}{rgb}{0.95,0.95,0.92}

\lstdefinestyle{mystyle}{
    backgroundcolor=\color{backcolour},   
    commentstyle=\color{codegreen},
    keywordstyle=\color{magenta},
    numberstyle=\tiny\color{codegray},
    stringstyle=\color{codepurple},
    basicstyle=\ttfamily\footnotesize,
    breakatwhitespace=false,         
    breaklines=true,                 
    captionpos=b,                    
    keepspaces=true,                 
    numbers=left,                    
    numbersep=5pt,                  
    showspaces=false,                
    showstringspaces=false,
    showtabs=false,                  
    tabsize=2
}

\graphicspath{ {./figures/} }
\makeatletter
\def\input@path{{./sections/}}
\makeatother

\newcommand{\eg}{{\it e.g.}}
\newcommand{\ie}{{\it i.e.}}
\newcommand{\etal}{{\it et al.}}
\newcommand{\sys}{pcapML\xspace}
\newcommand{\fe}{pcapML-FE\xspace}

\title{ Towards Reproducible Network Traffic Analysis }

\author{Jordan Holland}
\affiliation{%
  \institution{Princeton University}
  \country{}
}
\email{jordanah@princeton.edu}

\author{Paul Schmitt}
\affiliation{%
  \institution{USC/ISI}
  \country{}
}
\email{pschmitt@isi.edu}

\author{Prateek Mittal}
\affiliation{%
  \institution{Princeton University}
  \country{}
}
\email{pmittal@princeton.edu}

\author{Nick Feamster}
\affiliation{%
  \institution{University of Chicago}
  \country{}
}
\email{feamster@uchicago.edu}

\input{abstract}

\begin{document}
\sloppy

\maketitle
\pagestyle{plain}

\input{introduction}

\input{methodology}

\input{requirements}
\input{system}
\input{benchmarks}
\input{whereto}
\input{related}
\input{conclusion}\label{lastpage}

\microtypesetup{protrusion=false}

\bibliographystyle{ACM-Reference-Format}
\bibliography{paper}

\vfill
\pagebreak

\end{document}

%% file: sections/abstract.tex
\begin{abstract}
Analysis techniques are critical for gaining insight into network traffic given both the higher proportion of encrypted traffic and increasing data
rates. Unfortunately, the domain of network traffic analysis suffers from a lack 
of standardization, leading to incomparable results and barriers to 
reproducibility. Unlike other disciplines, no standard dataset format exists, 
forcing researchers and practitioners to create bespoke analysis pipelines 
for each individual task. Without standardization researchers cannot compare 
``apples-to-apples,'' preventing us from knowing with certainty if a new 
technique represents a methodological advancement or if it simply benefits
from a different interpretation of a given dataset.

In this work, we examine irreproducibility that arises from the lack of
standardization in network traffic analysis. First, we study the
literature, highlighting evidence of irreproducible research based on
different interpretations of popular public datasets. Next, we investigate
the underlying issues that have lead to the status quo and prevent reproducible
research. Third, we outline the standardization requirements that
any solution aiming to fix reproducibility issues must address. We then 
introduce \sys{}, an open source system which increases reproducibility of
network traffic analysis research by enabling metadata information to be
directly encoded into raw traffic captures in a generic manner. Finally, we 
use the standardization \sys{} provides to create the \sys{} benchmarks, 
an open source leaderboard website and repository built to track the progress of 
 network traffic analysis methods.
\end{abstract}

%% file: sections/introduction.tex
\section{Introduction}\label{sec:intro}

Researchers have developed methods to classify network traffic for over 30
years~\cite{mukherjee1994network}. Classification techniques have been used for
a variety of analysis tasks such as application detection, device
identification, intrusion detection, and website
fingerprinting~\cite{nmap,bernaille2006early,kdd98_dataset,panchenko2011website}.
In the past, techniques were able to leverage information inside of packet
payloads in order to easily classify traffic. However, with the rapid adoption
of encrypted network protocols, coupled with ever-increasing data volumes,
network traffic analysis has become both more important and more challenging.
Research in the field has largely turned to machine learning techniques to
address both of these challenges. Yet, after almost 20 years of applying
machine learning techniques to various traffic analysis problems, no standard
dataset format or comparison methodology exists~\cite{mcgregor2004flow}.

The lack of a standardized dataset format had led directly to a reproducibility
crisis in traffic analysis research: correctly reproducing previous work is
near-impossible. Researchers are required to build bespoke pipelines for each
new dataset in which they must engineer a pipeline to parse the format of the
dataset, organize the packets according to the task (\ie, applications,
intrusions, devices, websites), attach metadata to each set of separated
packets, and finally develop and evaluate an analysis technique. Testing even
the same technique on a different dataset requires re-inventing the wheel:
engineering a new pipeline for a new task. Worse, ambiguous terminology
commonly used throughout the field, such as a ``traffic flow'' or an
``application'' can lead to researchers, starting from the same dataset, to
test analysis techniques on completely different definitions of a task. 

\begin{figure*}[t]
    \centering
    \includegraphics[scale=0.25]{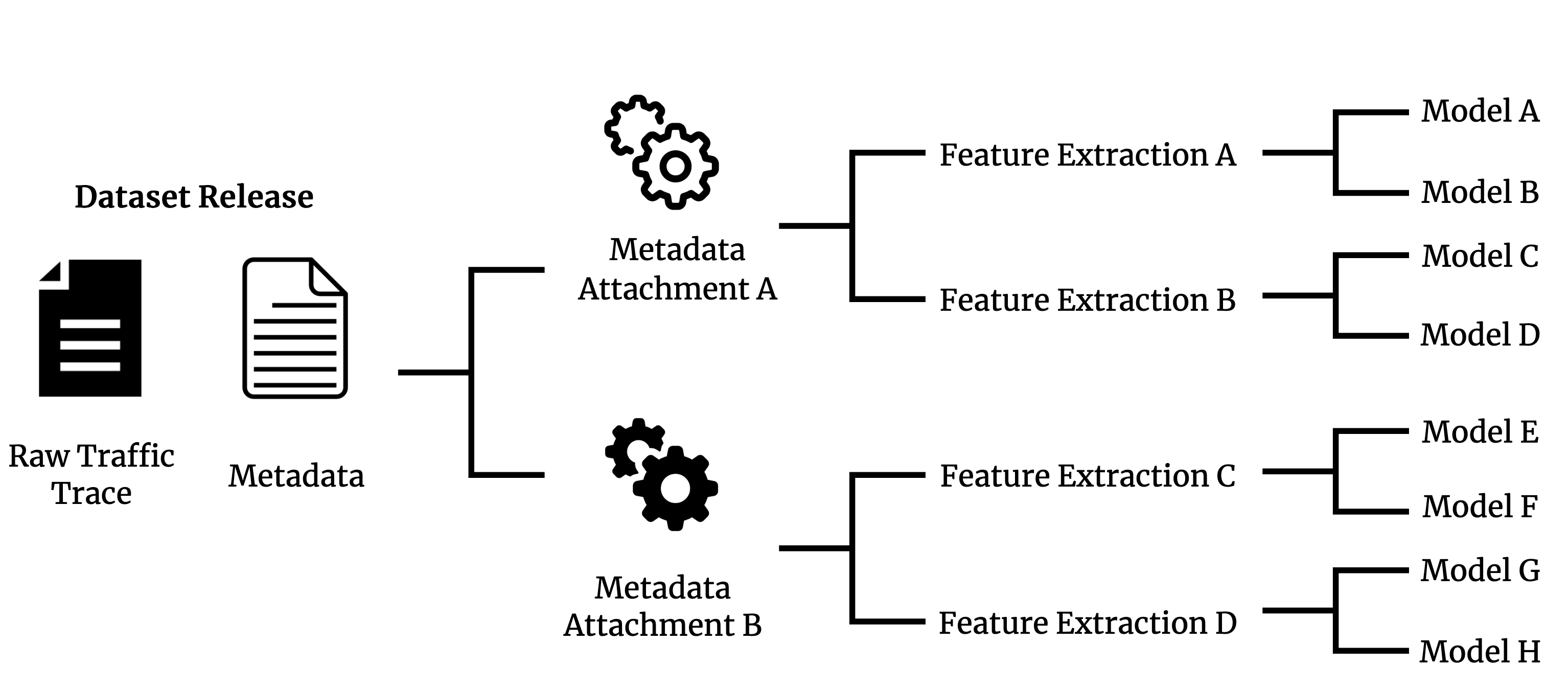}
    \caption{Releasing metadata separate from raw traffic leads to differing 
    versions of the same original task, 
    rendering it impossible to directly compare results.}
    \label{fig:branching}
\end{figure*}

These issues lead to multiple downstream problems. First, analysis techniques,
such as the performance of a machine learning classifier, are compared against
one another when the underlying problem definition differs. Second, the
difficulty of creating analysis pipelines has led to a focus on pre-processed
datasets, in which the dataset curators release features extracted from the
original network traffic as opposed to the raw packets. Techniques stemming
from these pre-processed datasets are inherently limited to a subset of the
released features and resulting research ultimately ends up being a ``model
bake-off,'' necessarily precluding the discovery of unforeseen features for a
given task. Finally, all of these issues create a barrier not only to
reproducibility, but to innovation in the field. When compared with the
progress of other fields, such as image recognition or natural language
processing, techniques in network traffic analysis have been relatively
stagnant despite increasing need for accurate analysis methods.

This work examines the state of network traffic analysis research. We first
highlight evidence of reproducibility challenges using examples from multiple
public datasets. We then highlight underlying causes of these reproducibility
issues, including the use of ambiguous terminology, the lack of a standardized
dataset format, and a focus on pre-processed datasets. Next, we take our
learnings and outline a list of requirements that any proposed
standardization solution must meet.

Finally, we introduce \sys{}, an open source system that aims to meet our
requirements and standardize network traffic analysis research. Rather than
focus on a standardized feature set, which previous work has called for, \sys{}
standardizes network traffic analysis research \textit{at the dataset
level}~\cite{barut2020netml,booij2021ton_iot}. \sys{} does this by enabling
researchers to encode metadata and traffic definitions (\ie, traffic flow
directionality) directly into raw traffic captures in a generic manner.
Further, \sys{}-encoded datasets can still be analyzed by the most popular
traffic analysis tools and libraries. Further, we release \fe{}, an open source
python library which lowers the bar for researchers to incorporate
\sys{}-encoded datasets into existing pipelines. Lastly, we use \sys{} to
create the \sys{} benchmarks a public leaderboard and repository for any 
traffic analysis task and dataset, enabling dataset creators to list their 
work on a central platform, dataset users to directly compare methods on a variety
of tasks, and the field to better track the progress of analysis techniques.

The rest of the paper organized as follows. Section~\ref{sec:need} presents
evidence of the need for standardization in network traffic analysis.
Section~\ref{sec:req} uses the lessons learned from Section~\ref{sec:need} to
outline a list of requirements for any standardization solution.
Section~\ref{sec:system} presents \sys{} and \fe{}, open source systems built
to increase reproducibility and increase innovation in the field. Next,
Section~\ref{sec:benchmark} presents the \sys{} benchmarks, an open platform
for dataset creators and users to develop and compare research. 
Section~\ref{sec:gen} then provides general recommendations for sound, 
reproducible research moving forward. Finally,
Sections~\ref{sec:related} and~\ref{sec:conclusion} examine related works and
summarize our contributions.

%% file: sections/methodology.tex
\section{The Need For Standardization}\label{sec:need}

\input{figures/tables/methodology/kdd.tex}

In this section we examine irreproducibility in the current
network traffic analysis ecosystem. First, we investigate literature leveraging
several popular network traffic datasets, finding that standard practices lead
to incomparable and irreproducible research, visualized in Figure~\ref{fig:branching}.
We then survey the literature to outline the practices that lead to irreproducible research.

\input{figures/tables/methodology/cicids.tex}

\subsection{Examples of reproducibility challenges.}
\label{subsec:reproc_example}

We begin by examining popular network traffic datasets and literature that
tests methods using those datasets. The following subsections highlight three
popular datasets examined. We describe how the datasets are curated, the
release format of the datasets, and finally compare literature leveraging the
datasets to demonstrate examples of irreproducibility in network
traffic analysis.

We wish to clearly state that the datasets and papers mentioned \textit{are not}
chosen to point out methodological errors by these specific dataset curators
or paper authors. On the contrary, the referenced datasets and papers generally
represent a \textit{higher} standard of reproducibility than many of the
datasets and works we examined in our search in that they outline their
methodology and dataset usage in such a way that comparisons can be made.

\subsubsection{DARPA 1998}

\paragraph{Overview.}
The DARPA 1998 intrusion detection dataset is perhaps the oldest and most well
known intrusion detection dataset~\cite{kdd98_dataset}. The dataset was created
in an effort to evaluate and compare intrusion detection methods on the same
task. In all, the dataset consists of seven weeks of training data and two
weeks of testing data, containing both benign network traffic and 38 attacks
comprising 4 broader classes of attack to be classified. Although this dataset
has known criticisms in terms of class distribution and experimental collection
it has still been widely used due to its long-term
availability~\cite{mchugh2000testing}.

\paragraph{Format.}
The network traffic was captured using \texttt{tcpdump} and released in the
PCAP file format~\cite{tcpdump}. As the challenge consisted of traffic across
ten weeks, a PCAP file was released for the traffic captured for each day of
the experiment. In all, this results in 45 separate PCAP files. The metadata
for the traffic was released in separate ``list'' files, which are CSV-like in
structure. For each PCAP file, a list file was released with labels for TCP and
UDP sessions found in the raw traffic. Each record in the metadata files lists
the start date, start time in Hour:Minute:Second format, the four-tuple (source
IP, destination IP, source port, destination port), and the label for the
session.

\paragraph{Evidence of Irreproducibility.}
Researchers using the dataset are required to build a pipeline that parses the
raw network traffic by session, associates each record in the metadata file
with the set of raw packets that define the session, and extracts information
from the session to classify it. To highlight how this methodology can lead to
reproducibility and comparability issues in practice, we take the ``list''
files released with the dataset and calculate the distribution of samples in
each of the five classes of traffic. We then examine work from the literature
that leveraged the dataset to compare this class distribution with.

Table~\ref{tab:kdd} shows the results of this experiment. We see that authors
ultimately experiment with analysis methods on different versions of the
dataset. The number of samples in each class of traffic differs both from the
originally released dataset and each other. As such, comparisons of the
intrusion detection methods in the papers listed and the originally published
methods are difficult, if not impossible, to perform. Further, recreating
these works is near-impossible as to do so one must recreate an entire bespoke
pipeline that lead to a differing version of the underlying dataset. These
issues partly occur due to the difficult and messy nature of merging raw
network traffic and metadata. Wang~\etal{} describe issues they encounter
during this process~\cite{wang2017hast}: 
\begin{displayquote}
``The traffic format of DARPA1998 is non-split pcap, which must be split into
multiple network flow files. In addition, the label files contain a few
problems, such as duplicated records and incorrect labels. For example, the
label file ``Test/Week2/Friday'' contains a record of ``07/32/1998,'' which is
an obvious date error. Therefore, the dataset requires preprocessing before the
experiments can be conducted.'' 
\end{displayquote}

\subsubsection{CICIDS 2017.}

\input{figures/tables/methodology/vpn-nonvpn.tex}

\paragraph{Overview.}
The CICIDS 2017 intrusion detection dataset was specifically curated to
represent a more modern intrusion detection dataset than previously
available~\cite{cicids_dataset}. The dataset consists of over 40GB of network
traffic captured over the course of five days. 15 different classes of network
traffic exist in the dataset: 14 types of attack traffic flows and benign
traffic flows.

\paragraph{Format.}
The dataset was released in two formats. First, a pre-processed set of features
in which the authors extract features from each labeled flow for use with
machine learning methods, using a custom built system. Second, the raw network
traffic and associated labels were released as a set of PCAP files, one to two
PCAP files per day of the experiment, and a metadata CSV containing information
to link the raw traffic flows to a label, such as the flow start timestamps and
flow 5-tuples.

\paragraph{Evidence of Irreproducibility.}
We again process the metadata files released with the dataset, determining the
class distribution of the dataset. We then examine the literature for work
testing methods on the dataset which reports the class distribution used for
their experiments. Table~\ref{tab:cicids} compares the class distribution in
the original dataset compared with work leveraging the dataset.

Table~\ref{tab:cicids} again shows that authors ultimately test methods on
different interpretations of the same dataset. No row in Table~\ref{tab:cicids}
is identical to another. Various works leveraging the dataset merge multiple
classes into a single class, subsequently downsampling the number of flows in
the merged class. Others ignore classes in the original dataset altogether.
Research developing and testing methods on differing interpretations of
the same dataset render it difficult to reason about the superiority of any
intrusion detection method over another.

\subsubsection{VPN-nonVPN}

\paragraph{Overview.}
The third dataset we examine in detail is the VPN-nonVPN dataset, released in
2016~\cite{draper2016characterization}. The dataset consists of  
traffic from 7 different types of applications, such as ``web browsing'' or
``streaming.'' The dataset contains both VPN and non-VPN traffic for each type
of application, consisting of 14 different labeled traffic classes.

\paragraph{Format.}
The dataset is released as a set of PCAP traffic files. Each PCAP file contains
traffic generated by an application. Metadata for the traffic is encoded in the
name of each PCAP, as each PCAP file is named by the specific application that
generated the traffic, such as \texttt{email1a.pcap}. No information mapping
the network traffic to the broader application type, which is ultimately used
for evaluation in the paper, is released. There is a description of the
specific applications that make up each ``application type'' in the paper
itself.

\input{figures/tables/methodology/terms.tex}

\paragraph{Evidence of Irreproducibility.}
We survey the literature leveraging this dataset to understand how it is used.
Although we are unable to gather detailed information regarding the class
distribution for each work, we do find the number of classes of traffic used
for evaluation in the original work and subsequent work.

As with previous examples, Table~\ref{tab:vpn} illustrates that works
leveraging the dataset interpret the dataset differently from one another.
Table~\ref{tab:vpn} also records the authors labeling process as described in
each paper, demonstrating how differing versions of the same dataset can be
created. Wang~\etal{} even note that they correspond
with the original authors and still cannot fully reproduce the original
dataset, ultimately testing methods on a dataset with two fewer classes than the
original work~\cite{wang2017end}.

\subsection{Causes of Irreproducibility}

\input{figures/tables/methodology/datasets.tex}

Subsection~\ref{subsec:reproc_example} presented examples of irreproducibility in 
network traffic analysis. In this section, we outline and investigate the causes 
of irreproducibility in network traffic analysis.

\subsubsection{Ambiguous Terminology}

The datasets examined in Subsection~\ref{subsec:reproc_example} describe 
network traffic and analysis tasks using terminology common throughout the
field. The DARPA 1998 analysis task focuses on ``sessions'', which correspond
to individual TCP or UDP ``connections'' between two IP
addresses. The CICIDS 2017 analysis task examines
``flows'', which they define as bi-directional using the 5-tuple definition of
a flow (source IP, destination IP, source port, destination port, protocol).
Finally, the VPN-nonVPN dataset focuses on ``application'' identification,
where applications are defined by single traffic flow which is considered to be
bi-directional.

Although vernacular is useful in conversation, precise definitions of these
terms are required for reproducible research. We perform a simple experiment to
better understand the varied definitions of similar terminology in traffic
analysis. We examine work appearing in ACM CCS, USENIX Security, PETS
Symposium, IEEE S\&P, and NDSS in the years 2021, 2020, and 2019, further
supplementing our search with papers already included in
Subsection~\ref{subsec:reproc_example}. We search each paper for two types of
terms. First, we look for the most specific term used for defining the analysis
task at hand, such as ``5-tuple traffic flow'' or ``application''. Second, we
examine terms used when defining features for classification, such as
``sub-flow'' or ``tcp-stream''. We do not simply report if a term appeared in a
paper in any capacity, nor do we consider subsets of terms (\ie, 5-tuple flow
does not also include flow). We examine and extract terminology used
in 50 papers in total.

Table~\ref{tab:terminology} shows the 25 separate terms found during the
search. Perhaps more concerning than the raw number of terms used to describe
analysis tasks is that even terms \textit{in the same row} of
Table~\ref{tab:terminology} can vary in definition. For example, works can
consider 5-tuple flows as either uni-directional or
bi-directional~\cite{cicids_dataset,zhang2019network}. We further detail two
classes of terminology below and describe how ambiguous terminology can
directly lead to researchers working on non-identical tasks.

\paragraph{Flows.}
The term ``flow'' is present throughout a large number of papers we examine,
but defined differently in many of the papers. For example, some works define a
flow using a 4-tuple (source IP, destination IP, source port, destination
port), while others use a 5-tuple definition (4-tuple and protocol). Worse,
each of these definitions can be further divided into uni-directional and
bi-directional versions. These distinctions can, and have, lead to the
branching methodologies shown in Figure~\ref{fig:branching}. For
example, Sharafaldin~\etal{} curated, released, and provided an intrusion
detection method for the CICIDS dataset~\cite{cicids_dataset}. In their work,
they analyze the dataset considering flows as bi-directional 5-tuples, while a
5-tuple flow can also be thought of as uni-directional. Zhang~\etal{}
subsequently leverage the raw CICIDS dataset but use software to parse 
the traffic which considers 5-tuple flows as uni-directional, 
representing a different task than the original authors.
Combining this issue with the varying class distributions shown in
Table~\ref{tab:cicids} only creates more methodology branches which are not
directly comparable with others. 

Even small modifications to the definition of units of traffic can directly
affect the results of analysis methods. Draper-Gil~\etal{} examine this,
analyzing the performance of models trained on an application identification
task using four different flow timeout values, finding that the accuracy of the
models varied by up to 3\%~\cite{draper2016characterization}. In another
example, Garcia~\etal{} redefine and re-release a popular public dataset based
off of bi-directional traffic flows outperforming a uni-directional flow
definition on the same task~\cite{garcia2014empirical}:
\begin{displayquote}
``Each scenario was captured in a pcap file that contains all the packets of the 
three types of traffic. These pcap files were processed to obtain other type of 
information, such as NetFlows, WebLogs, etc. The first analysis of the 
CTU-13 dataset, that was described and published in the paper 
"An empirical comparison of botnet detection methods" (see Citation below) 
used unidirectional NetFlows to represent the traffic and to assign the labels. 
These unidirectional NetFlows should not be used because they were outperformed 
by our second analysis of the dataset, which used bidirectional NetFlows.''
\end{displayquote}.

\paragraph{Applications and Websites.}
The ``application identification'' and ``website fingerprinting'' tasks further
exhibit the perils of ambiguous terminology in traffic analysis. Clearly
defining the set of packets that are attributable to a single application or
website is critical to comparing, contextualizing, and reproducing methods for
a given task, as definitions of these tasks can drift over time. In the past,
applications often used a single network flow (5-tuple,
uni-directional)~\cite{bernaille2006early}. However, modern applications
commonly leverage multiple simultaneous traffic flows (\eg, Netflix video
streaming can include many several underlying flows for a single video
session), and recent research has used this information to analyze
traffic~\cite{bronzino2019inferring}. Website fingerprinting tasks raise
similar questions: does a website trace comprise of only traffic to and from a
single server IP address, or are the multiple connections to other server IPs
considered when performing analysis? 

\subsubsection{Different Tasks, Different Dataset Formats.}

The three datasets presented in Subsection~\ref{subsec:reproc_example} comprise
two discrete tasks: application classification and intrusion detection. Yet,
each of the three datasets requires a custom-built analysis pipeline to develop
and test methods. Although all three datasets store their network traffic in
the PCAP capture format, the metadata, which defines how to separate the
packets into an ``intrusion'' or an ``application'', is released in a variety
of formats. Table~\ref{tab:datasets}, which shows the release format of a
variety of popular datasets, demonstrates that this is common practice:
datasets release network traffic in PCAP format and metadata in a one-off
manner. Worse, Table~\ref{tab:datasets} demonstrates that even for researchers
working only on a \textit{single task} (\ie, intrusion detection), there is no
common dataset format to interact with. Each dataset format increases the
engineering burden on researchers developing new methods, increasing the
chances for human error and resulting in branching methodologies shown in
Figure~\ref{fig:branching}.

\subsection{Preprocessed Datasets Inhibit Innovation.}

Another trend seen in Table~\ref{tab:datasets} and subsequent research is a
heavy focus on releasing pre-processed traffic as well as calls for
standardized feature sets~\cite{barut2020netml,booij2021ton_iot}. The highest
cited dataset in Table~\ref{tab:datasets}, the NSL-KDD dataset, is a set of
pre-processed features derived from the KDD99 dataset which was originally
derived from the DARPA1998 and DARPA1999 datasets\cite{kayacik2005selecting}. 

The predictive accuracy of any machine learning algorithm, regardless of task,
is impacted by the representation of the data that is offered to the model as
input. Typically, machine learning pipelines are developed iteratively,
observing model performance using different representations of the raw data. It
stands to reason that it can be difficult, if not impossible, to know which
features should be extracted and how data is best represented for any given
task {\em a priori}. Publicly released datasets of pre-computed features,
rather than raw network traffic that was used to generate the features,
prohibit future researchers entirely from exploring new features that the
original work may not have considered. This practice directly hinders advances
in \textit{techniques}, forcing research based on pre-processed datasets to
ultimately boil down to model ``bake-offs.''

%% file: figures/tables/methodology/kdd.tex
\begin{table}[t]
    \centering
 \resizebox{\columnwidth}{!}{
\begin{tabular}{@{}llllllll@{}}
\cmidrule(l){4-8}
\multicolumn{1}{c}{}       & \multicolumn{1}{c}{}          & \multicolumn{1}{c}{}     & \multicolumn{5}{c}{Class Distribution}                                                                                               \\ \midrule
\multicolumn{1}{c}{Source} & \multicolumn{1}{c}{Citations} & \multicolumn{1}{c}{Year} & \multicolumn{1}{c}{Normal} & \multicolumn{1}{c}{DoS} & \multicolumn{1}{c}{Probe} & \multicolumn{1}{c}{R2L} & \multicolumn{1}{c}{U2R} \\ \midrule
\rowcolor[HTML]{C0C0C0} 
    Dataset Release~\cite{kdd98_dataset}            & 1287                          & 2000                     & 1,157,873                  & 1,919,937               & 54,793                    & 949                     & 48                      \\
    Khan~\etal{}~\cite{khan2007new}               & 498                           & 2007                     & 878,318                    & 308,808                 & 15,120                    & 3,327                   & 39                      \\ 
\rowcolor[HTML]{C0C0C0} 
    Wang~\etal{}~\cite{wang2017hast}               & 301                           & 2017                     & 1,309,598                  & 2,152,850               & 89,301                    & 14,535                  & 436                     \\ \bottomrule
\end{tabular}
    }%
\caption{Usage of the KDD-CUP98 dataset differs from both the original release and other works.}
\label{tab:kdd}
\end{table}

%% file: figures/tables/methodology/cicids.tex
\begin{table*}[t]
    \centering
 \resizebox{\textwidth}{!}{
\begin{tabular}{lrrrrrcrrrrrrrrcrr}
\hline
                                        & \multicolumn{1}{l}{}          & \multicolumn{16}{c}{Class Distribution}                                                                                                                                                                                                                                                                                                                                                                                                                                                                                                                                                               \\ \cline{4-18} 
\multicolumn{1}{c}{}                    & \multicolumn{1}{c}{}          & \multicolumn{1}{c}{}     & \multicolumn{1}{c}{}       & \multicolumn{1}{c}{}     & \multicolumn{1}{c}{}                                & \multicolumn{4}{c}{DoS}                                                                                                                          & \multicolumn{1}{c}{}           & \multicolumn{1}{c}{}             & \multicolumn{1}{c}{}         & \multicolumn{1}{c}{}            & \multicolumn{1}{c}{}                               & \multicolumn{3}{c}{Web}                                                                                         \\ \cline{6-10} \cline{16-18} 
\multicolumn{1}{c}{Source}              & \multicolumn{1}{c}{Citations} & \multicolumn{1}{c}{Year} & \multicolumn{1}{c}{Benign} & \multicolumn{1}{c}{Bot}  & \multicolumn{1}{c}{DDoS}                            & GoldenEye                                          & \multicolumn{1}{c}{Hulk} & \multicolumn{1}{c}{Slowhttptest} & \multicolumn{1}{c}{Slowloris} & \multicolumn{1}{c}{Heartbleed} & \multicolumn{1}{c}{Infiltration} & \multicolumn{1}{c}{PortScan} & \multicolumn{1}{c}{FTP-Patator} & \multicolumn{1}{c}{SSH-Patator}                    & Brute Force                                       & \multicolumn{1}{c}{SQL Injection} & \multicolumn{1}{c}{XSS} \\ \hline
\rowcolor[HTML]{C0C0C0} 
    {\color[HTML]{333333} Original Dataset~\cite{cicids_dataset}} & 1,057                         & 2018                     & 2,273,097                  & 1,966                    & 128,027                                             & \multicolumn{1}{r}{\cellcolor[HTML]{C0C0C0}10,293} & 231,073                  & 5,499                            & 5,796                         & 11                             & 36                               & 158,930                      & 3,938                           & 5,879                                              & \multicolumn{1}{r}{\cellcolor[HTML]{C0C0C0}1,507} & 21                                & 652                     \\
    Vinakayumar~\etal{}~\cite{vinayakumar2019deep}                     & 484                           & 2019                     & 80,000                     &
    1,966                    & \multicolumn{1}{r|}{8,000}                          & \multicolumn{4}{c|}{8,000 $\blacklozenge$}
    & -                              & -                                & 8,000                        & 7,938                           &
    \multicolumn{1}{r|}{5,879}                         & \multicolumn{3}{c|}{2,180$\blacklozenge$}                                                                                      \\
\rowcolor[HTML]{C0C0C0} 
    Zhang~\etal{}~\cite{zhang2019network}                           & 63                            & 2019                     & 339,621                    & 1,441                    & 16,050                                              & \multicolumn{1}{r}{\cellcolor[HTML]{C0C0C0}7,458}  & 14,108                   & 4,216                            & 3,869                         & -                              & -                                & 158,673                      & 3,907                           & 2,511                                              & \multicolumn{1}{r}{\cellcolor[HTML]{C0C0C0}1,353} & 12                                & 631                     \\
    Zhou et. al~\etal{}~\cite{zhou2020building}                             & 134                           & 2020                     & 439,683                    & -                        & -                                                   & \multicolumn{1}{r}{10,293}                         & 230,124                  & 5,499                            & 5,796                         & 11                             & -                                & -                            & -                               & -                                                  & \multicolumn{1}{r}{-}                             & -                                 & -                       \\
\rowcolor[HTML]{C0C0C0} 
    Barut~\etal{}~\cite{barut2020netml}                            & 5                             & 2020                     & 248,067                    & -
    & \multicolumn{1}{r|}{\cellcolor[HTML]{C0C0C0}45,168} & \multicolumn{4}{c|}{\cellcolor[HTML]{C0C0C0}29,754$\blacklozenge$}
    & -                              & 66,914                           & 153,028                      & 3,958                           &
    \multicolumn{1}{r|}{\cellcolor[HTML]{C0C0C0}2,464} & \multicolumn{3}{c|}{\cellcolor[HTML]{C0C0C0}2,019$\blacklozenge$}                                                              \\
    Stiawan~\etal{}~\cite{stiawan2020cicids}                          & 47                            & 2020                     & 454,396                    &
    \multicolumn{1}{r|}{367} & \multicolumn{5}{c|}{76,265$\blacklozenge$}                                                                                                                                                                            & -                              & 6                                & 32,882                       & 2,717                           & -                                                  & \multicolumn{1}{r}{426}                           & -                                 & -                       \\ \hline
\end{tabular}
} %
    \caption{Work stemming from the same dataset ends up testing methods on differing versions of the same dataset, rendering direct comparisons impossible.
    ($\blacklozenge$ denotes merged classes.)}
\label{tab:cicids}
\end{table*}

%% file: figures/tables/methodology/vpn-nonvpn.tex
\begin{table*}[t]
    \centering
 \resizebox{\textwidth}{!}{
\begin{tabular}{@{}lrrl@{}}
\toprule
Paper             & \multicolumn{1}{l}{Citaitons} & \multicolumn{1}{l}{Classes} & Labeling Description                                                                                                                                                                                                                                                                                                                                                                                                                                                                                                                                                                                                                                                                                                                                                                                                                                                                                                                           \\ \midrule
\rowcolor[HTML]{C0C0C0} 
    Draper-Gil~\etal{}~\cite{draper2016characterization} & 365                           & 14                          & Original Dataset                                                                                                                                                                                                                                                                                                                                                                                                                                                                                                                                                                                                                                                                                                                                                                                                                                                                                                                               \\
    Wang~\etal{}~\cite{wang2017end}       & 346                           & 12                          & \begin{tabular}[c]{@{}l@{}}``The flow features of ISCX
    dataset have 14 classes of labels, but the raw traffic has no labels, so we labeled pcap files in the dataset according to the description of their paper.\\
    Some files such as “Facebook\_video.pcap” can be labeled as either “Browser” or “Streaming”, and all files related to “Browser” and “VPN-Browser” have this
    problem. \\ We can’t solve this problem even after email communication with the authors, so we decided not to label these files.''\end{tabular}                                                                                                                                                                                                                                                                                                                                                                                                                  \\
\rowcolor[HTML]{C0C0C0} 
Lotfollahi~\etal{}~\cite{lotfollahi2020deep} & 385                           & 12, 17                      & \begin{tabular}[c]{@{}l@{}}``..the dataset’s pcap
files are labeled according to the applications and activities they were engaged in. However, for application identification and traffic characterization tasks,
\\ we need to redefine the labels, concerning each task. For application identification, all pcap files labeled as a particular application which were collected
during a nonVPN session, \\ are aggregated into a single file. This leads to 17 distinct labels shown in Table 1a. Also for traffic characterization, we
aggregated the captured traffic of different applications involved \\ in the same activity, taking into account the VPN or non-VPN condition, into a single pcap
file. This leads to a 12-classes dataset, as shown in Table 1b.''\end{tabular}                                                                                                                                                     \\
Zou~\etal{}~\cite{zou2018encrypted}        & 41                            & 12                          & \begin{tabular}[c]{@{}l@{}}``The dataset contains
25GB raw traffic in the pcap format, which includes 14 network application classes, where 7 for regular traffic (such as Spotify  and Facebook) and the rest for
the \\ corresponding traffic with VPN encrypted (such as VPN- Spotify and VPN-Facebook). We relabel the raw traffic into 12 classes, to be used in our
experiments...''\end{tabular}                                                                                                                                                                                                                                                                                                                                                                                                                                                                                                                                          \\
\rowcolor[HTML]{C0C0C0} 
Zeng~\etal{}~\cite{zeng2019deep}      & 96                            & 7                           & \begin{tabular}[c]{@{}l@{}}``The first selected dataset is
regenerated from ISCX VPN-nonVPN traffic dataset in order to evaluate the effectiveness of DFR on encrypted traffic classification. ISCX VPN-nonVPN dataset \\
originally has 7 types of regular encrypted traffic and 7 types of protocol encapsulated traffic. Since we mainly focus on evaluating the efficiency on
encrypted traffic classification, we will select \\ and label data from those 7 types of regular encrypted traffic, which are Web Browsing, Email, Chat,
Streaming, File Transfer, VoIP, and P2P. To be noticed that all other six types of encrypted \\ traffic are related to Web Browsing, hence we abandoned his
class of encrypted traffic referring to Wang’s work.''\end{tabular}                                                                                                                                                                       \\
Barut~\etal{}~\cite{barut2020netml}     & 5                             & 7, 18, 31                   & \begin{tabular}[c]{@{}l@{}}``The third dataset focuses
on application type classification and is called non-vpn2016. It is obtained by extracting flow features using only the non-vpn raw traffic capture files from
the \\ CIC website as our feature extraction tool does not support VLAN processing yet. Three levels of annotations are assigned to this dataset: top-level,
mid-level and fine-grained. \\ Top-level annotations are a general grouping of those traffic capture data and 7 classes are selected including P2P, audio, chat,
email, file\_transfer, tor, video. \\ Mid-level annotations contain 18 type of applications  (facebook, skype etc.) while fine-grained annotations identify 31
lower-level classes in an application \\ (facebook\_audio, facebook\_chat, skype\_audio, skype\_chat\textbackslash{}\textbackslash etc.).  Table 4 gives the
list of files used to create non-vpn2016 dataset.''\end{tabular} \\
\rowcolor[HTML]{C0C0C0} 
Wang~\etal{}~\cite{wang2018datanet}       & 106                           & 15                          & \begin{tabular}[c]{@{}l@{}}``The dataset for
evaluation is selected from the ‘‘ISCX VPN-nonVPN traffic dataset’. As shown in Table 2 the total dataset for evaluation is composed of 15 applications, e.g.,
Facebook, \\ Youtube, Netflix, etc. The chosen applications are encrypted with various security protocols, including HTTPS, SSL, SSH, and proprietary protocols.
A total of 206,688 data packets \\ are included in the selected dataset.''\end{tabular}                                                                                                                                                                                                                                                                                                                                                                                                                                                                             \\ \bottomrule
\end{tabular}
    }%
\caption{The current traffic analysis ecosystem creates a burden for dataset curators and researchers using the dataset. Dataset curators are forced to choose a format to release the dataset and work with researchers to rebuild the dataset. Researchers using the dataset must
attempt to recreate the datasets faithfully.}
\label{tab:vpn}
\end{table*}

%% file: figures/tables/methodology/terms.tex
\begin{table}[t]
    \centering
    \small
\begin{tabular}{@{}ll@{}}
\toprule
Term                   & Referenced By                                                                                                                                                                                                \\ \midrule
\rowcolor[HTML]{C0C0C0} 
5-tuple flow           & \cite{barradas2021flowlens,sharafaldin2018toward,zhang2019network,zhou2020building,barut2020netml,wang2017end,zou2018encrypted,mazhar2018real,gutterman2019requet,nprint} \\
4-tuple flow           & \cite{tolley2021blind,van2020flowprint,frolov2019use}                                                                                                                                  \\
\rowcolor[HTML]{C0C0C0} 
application            & \cite{barut2020netml,nprint}                                                                                                    \\
bi-directional flow    & \cite{draper2016characterization,barut2020netml,zeng2019deep}                                                                                                                            \\
\rowcolor[HTML]{C0C0C0} 
bi-directional session & \cite{wang2017end}                                                                                                                                                                                           \\
channel                & \cite{barradas2018effective,mirsky2018kitsune}                                                                                                                                                               \\
\rowcolor[HTML]{C0C0C0} 
dns encrypted flow     & \cite{siby2019encrypted}                                                                                                                                                                                     \\
flow                   & \cite{van2020flowprint,bock2019geneva,apthorpe2018keeping,barut2020netml,lotfollahi2020deep}                                                                                                  \\
\rowcolor[HTML]{C0C0C0} 
network flow           & \cite{nasr2018deepcorr,vinayakumar2019deep,kdd98_dataset}                                                                                                                                                    \\
network trace          & \cite{qasem2019finding}                                                                                                                                                            \\
\rowcolor[HTML]{C0C0C0} 
packet trace           & \cite{li2018measuring}                                                                                                                                                                                       \\
stream                 & \cite{jansen2018inside}                                                                                                                                                                                      \\
\rowcolor[HTML]{C0C0C0} 
sub-flow               & \cite{sharafaldin2018toward,zhou2020building,stiawan2020cicids}                                                                                                                                              \\
tcp connection         & \cite{bock2021weaponizing,wang2021themis,bock2019geneva,alexander2019detecting,kdd98_dataset}                                                                                             \\
\rowcolor[HTML]{C0C0C0} 
tcp flow               & \cite{jero2018automated}                                                                                                                                                                      \\
tcp session            & \cite{wang2020symtcp,antonioli2019nearby,kdd98_dataset}                                                                                                                                                      \\
\rowcolor[HTML]{C0C0C0} 
tcp stream             & \cite{jansen2021once,wang2020symtcp,bock2019geneva,kohls2019challenges}                                                                                                                                      \\
tcp sub-flow           & \cite{yan2018feature}                                                                                                                                                                                        \\
\rowcolor[HTML]{C0C0C0} 
traffic burst          & \cite{zhang2018homonit}                                                                                                                                                                                      \\
traffic flow           & \cite{zhang2018homonit,barradas2018effective,mazhar2018real,nprint}                                                                                                                                          \\
\rowcolor[HTML]{C0C0C0} 
traffic stream         & \cite{mazhar2018real}                                                                                                                                                                                        \\
traffic trace          & \cite{apthorpe2018keeping,sirinam2018deep,rimmer2017automated,jansen2018inside}                                                                                                                              \\
\rowcolor[HTML]{C0C0C0} 
udp connection         & \cite{kdd98_dataset}                                                                                                                                                                                         \\
udp flow               & \cite{van2020flowprint}                                                                                                                                                                   \\
\rowcolor[HTML]{C0C0C0} 
website trace          & \cite{smith2021website,oh2021gandalf,gong2020zero,rahman2019tik,sirinam2019triplet,oh2017p,bhat2019var,li2018measuring,rimmer2017automated}                                                                  \\ \bottomrule
\end{tabular}
\caption{Terminology varies among work examined.}
\label{tab:terminology}
\end{table}

%% file: figures/tables/methodology/datasets.tex
\begin{table*}[t]
    \centering
 \resizebox{\textwidth}{!}{
\begin{tabular}{@{}llllll@{}}
\toprule
    \multicolumn{1}{c}{Dataset}                   & \multicolumn{1}{c}{Citations} & \multicolumn{1}{c}{Task}   & \multicolumn{1}{c}{Raw Traffic (PCAP)?} & \multicolumn{1}{c}{\begin{tabular}[c]{@{}c@{}}Metadata \\ Format\end{tabular}} & \multicolumn{1}{c}{\begin{tabular}[c]{@{}c@{}}Preprocessed \\ Features?\end{tabular}} \\ \midrule
\rowcolor[HTML]{C0C0C0} 
DARPA 1998~\cite{kdd98_dataset}               & 1,288                         & Intrusion Detection        & \CheckedBox               & List File                                                                      &                                                                                       \\
DARPA 1999~\cite{lippmann20001999}            & 1,250                         & Intrusion Detection        & \CheckedBox               & Described On Website                                                           &                                                                                       \\
\rowcolor[HTML]{C0C0C0} 
UNSW-NB15~\cite{moustafa2015unsw}             & 1179                          & Intrusion Detection        & \CheckedBox               & ARGUS, BRO, CSV                                                                & \CheckedBox                                                                           \\
TON\_IoT~\cite{alsaedi2020ton_iot}             & 60                            & Cybersecurity Applications & \CheckedBox               & LOG, CSV                                                                       & \CheckedBox                                                                           \\
\rowcolor[HTML]{C0C0C0} 
Bot-IoT~\cite{koroniotis2019towards}          & 376                           & Botnet Detection           & \CheckedBox               & ARGUS, BRO, CSV                                                                & \CheckedBox                                                                           \\
CICIDS 2017~\cite{cicids_dataset}             & 1,057                         & Intrusion Detection        & \CheckedBox               & CSV                                                                            & \CheckedBox                                                                           \\
\rowcolor[HTML]{C0C0C0} 
VPN-nonVPN~\cite{draper2016characterization}  & 365                           & Application Type           & \CheckedBox               & Described In paper                                                             & \CheckedBox                                                                           \\
Android Malware 2017~\cite{shiravi2012toward} & 938                           & Malware Detection          & \CheckedBox               & CSV                                                                            & \CheckedBox                                                                           \\
\rowcolor[HTML]{C0C0C0} 
Tor 2016~\cite{lashkari2017characterization}  & 300                           & Application Type           & \CheckedBox               & Described On Website                                                           & \CheckedBox                                                                           \\
CTU-13~\cite{garcia2014empirical}             & 540                           & Botnet Detection           &                           & BIARGUS                                                                        & \CheckedBox                                                                           \\
\rowcolor[HTML]{C0C0C0} 
NSL-KDD~\cite{kayacik2005selecting}           & 3,339                         & Intrusion Detection        &                           & CSV                                                                            & \CheckedBox                                                                           \\
KDD99~\cite{kdd99}                            & -                             & Intrusion Detection        &                           & CSV                                                                            & \CheckedBox                                                                           \\
\rowcolor[HTML]{C0C0C0} 
Deep Fingerprinting~\cite{sirinam2018deep}    & 161                           & Website Fingerprinting     &                           & PKL                                                                            & \CheckedBox                                                                           \\ \bottomrule
\end{tabular}
}
\caption{Dataset formats and separate tasks create a large barrier to testing new methods}
\label{tab:datasets}
\end{table*}

%% file: sections/requirements.tex
\section{Standardization Requirements}\label{sec:req}

In this section we outline a list of requirements that a system for
standardized traffic analysis must meet. These requirements arise from issues
uncovered in Section~\ref{sec:need}.

\paragraph{Compatible.}
A large ecosystem of mature and well-known tools and libraries exists for
parsing, filtering, and analyzing network traffic, including \texttt{tcpdump},
\texttt{tshark}, \texttt{wireshark}, \texttt{pf\_ring}, and
\texttt{libpcap}~\cite{libpcap,tcpdump,wireshark,tshark,pfring}. Any solution
that is not inherently compatible with this long-developed ecosystem of tools
would amount to re-inventing the wheel and be highly unlikely to be adopted.

\paragraph{Standard.}
Section~\ref{sec:need} explored the pitfalls of requiring researchers to build
analysis pipelines for a variety of dataset formats. Any system for  
standardized traffic analysis must standardize all traffic analysis tasks and
datasets under a single dataset format. This requirement significantly lowers
the burden on researchers creating and reproducing traffic analysis research,
decreasing the chance for human error.

\paragraph{Portable.}
The standardization solution must \textit{unify} metadata and raw traffic
traces. This requirement eliminates engineering errors that can occur when
stitching together metadata and raw network traffic traces. Ultimately, any
standardization solution should enable dataset curators to release a single
file that contains all raw traffic and associated metadata whenever feasible. 

\paragraph{Unambiguous.}
Section~\ref{sec:need} examined the variety of terms that can be used to
categorize and identify network traffic, such as traffic flows, traffic
sessions, application traffic, and website traces. Vague terminology leads to
work that is incomparable. As such, a standardization solution must provide a
method for eliminating all ambiguity by encoding this information with the
network traffic.

\paragraph{Unprocessed.}
Pre-processed datasets make the assumption that the pre-processed features best
describe the traffic, reducing future methods to only a subset of the
preprocessed features. A standardization solution must occur at the lowest
granularity possible, the packet level, to avoid any loss of information that
could be leveraged in future work.

\paragraph{Reproducible.}
Any standardization solution must provide users with the ability to uniquely
identify all traffic samples in a dataset (\ie, traffic flows, applications,
devices). This requirement enhances the reproducibility of traffic analysis
work, by enabling researchers to directly compare new methods with previous
ones.

%% file: sections/system.tex
\section{\sys{}}\label{sec:system}


In this section we present \sys{} and \fe{} (Feature Explorer), open source
systems designed to address issues with current practices
highlighted in Section~\ref{sec:need} while meeting the requirements presented
in Section~\ref{sec:req}. We begin by discussing our solution for the \sys{}
output file format and embedded metadata. We then present the processing
pipeline of \sys{} using an example case study. Next, we evaluate costs
associated with \sys{} using multiple data sets. Finally, we present \fe{},
which eases the burden of incorporating \sys{} outputs into existing ML
pipelines. 

\begin{figure*}[t]
    \centering
    \includegraphics[scale=0.45]{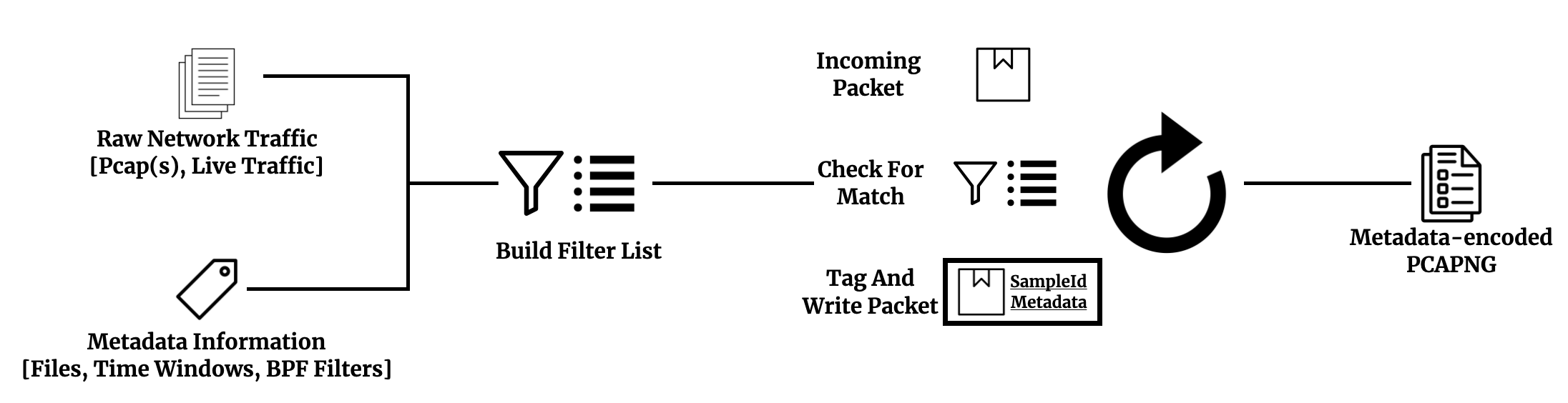}
    \caption{\sys{} enables researchers to directly encode metadata into raw traffic captures
    in a manner that familiar tools used to filter, read, and analyze network traffic can still be leveraged.}
    \label{fig:pcapml}
\end{figure*}

\subsection{Output Format: PCAPNG}

As outlined in Section~\ref{sec:req}, any standardization solution must meet
multiple requirements, including providing a standard format that is compatible
with the existing ecosystem of tools.

Almost universally, the networking community has settled on using the PCAP
capture format for capturing, storing, analyzing, filtering, and releasing
network traffic~\cite{pcap}. The PCAP file format is supported by every popular
traffic analysis tool, including \texttt{tcpdump}, \texttt{libpcap} and
\texttt{wireshark}~\cite{libpcap,tcpdump,wireshark}. Unfortunately, the PCAP
format, which can be conceptualized as a linked list of raw network packets,
provides no generalizable method to unify metadata and raw packets. Techniques
to embed metadata information into packet headers, such as optional IP or TCP
options, cannot be relied upon as there is no guarantee that these headers will
exist in every packet, or that such headers contain the necessary space for
metadata. This limitation has led to the status quo for datasets: traffic
captures and metadata are stored and released as separate files.

More recently (beginning in 2014), working groups have developed the PCAP next
generation (PCAPNG) traffic capture format~\cite{pcapng}. While the PCAP format
represents a linked list of raw packets, the PCAPNG file format can be
conceptualized as a linked list of blocks which encapsulate raw packets. Each
packet block contains the raw packet data, the timestamp the packet was
captured, and importantly, has a variable length \texttt{options} field. The
options field resembles other variable length option fields, such as the TCP
options in that it is represented as a list of code, value pairs. \sys{}
leverages the \texttt{options} field on each packet block to encode metadata
directly into the PCAPNG file, unifying metadata and raw network traffic. Our
use of PCAPNG satisfies several of our requirements presented in
Section~\ref{sec:req}, as discussed below.

\paragraph{Compatible.}
The PCAPNG file format has the benefit of \textit{already} being supported by
the most popular libraries and tools to capture and analyze network traffic.
\texttt{Wireshark} and \texttt{tshark} are able to both parse and output PCAPNG
traffic captures by default, and \texttt{libpcap} and \texttt{tcpdump} can
parse PCAPNG captures\cite{tcpdump,tsharkpcapng}. 

\paragraph{Standardized.}
PCAPNG's options field provides a means to unify metadata and raw network
traffic and create a standard, unified dataset format for holding both network
traffic and metadata. There are two options to attach metadata to each packet.
First, the PCAPNG file format contains a built-in option type,
\texttt{opt\_comment}, which allows for arbitrary UTF-8 strings to be attached
to every packet in a traffic capture. The advantage of using the built-in
option is the guaranteed portability of the encoding. The disadvantage of this
method is there is no specified structure for the metadata-encoding as the
comment is an arbitrary UTF-8 string. The second option for encoding metadata
into the packet block options field is to use a custom option type. This method
has the advantage of building structure into the option itself. Unfortunately,
the drawbacks include lack of guaranteed portability and flexibility in the
future. As such, \sys{} leverages the preexisting \texttt{opt\_comment} for
unifying metadata and raw traffic, using a generic CSV structure inside of the
UTF-8 string itself to encode information.

\paragraph{Portable.}
We leverage the PCAPNG format to directly embed metadata information for every
packet in a network traffic trace. As such, \sys{} creates a single, portable
solution for network traffic datasets, enabling researchers to build analysis
pipelines for multiple datasets and tasks around a single dataset format.

\paragraph{Unprocessed.}
\sys{} enables feature exploration by providing a solution at the finest
granularity possible, the packet level. Attaching metadata to each individual
packet in a dataset incurs no loss of information, allowing researchers to
explore, test, and compare potentially unseen features and methods for
analyzing the traffic with existing techniques.

\subsection{Metadata Format: SampleIDs} 

As discussed, the PCAPNG file format enables us to directly couple metadata and
individual network packets. We now require a generalizable method to match any
given packet with its associated metadata and traffic sample. For each traffic
analysis task, a traffic sample can be defined differently. In a traffic flow
identification task, all packets that belong to a given traffic flow comprise a
single traffic sample. For an OS detection task, a single packet may constitute
a traffic sample. For anomaly detection, all of the packets inside of a
specific time window can comprise a single traffic sample. Finally, for an
application identification task, all of the packets created by an application,
which may include multiple traffic flows, comprise a single traffic sample. 

The variety of goals and traffic sample definitions has led to bespoke
pipelines for each individual task. With \sys{}, our goal is to find a
generalizable solution for the field to build upon. We notice that all traffic
analysis tasks have an important common factor: \textit{every traffic sample
can be defined by a group of one or more packets}. We leverage this insight and
the ability to encode metadata into PCAPNG traffic captures to create
sampleIDs, a simple, generalizable method to identify traffic samples. \sys{}'s
use of sampleIDs fulfill two requirements that outlined in
Section~\ref{sec:req}.

\paragraph{Unambiguous.}
\sys{} eliminates ambiguity by generating a sampleID for each group of packets
that represents a traffic sample for a given analysis task. For example, \sys{}
can generate a sampleID for every packet belonging to the same traffic flow,
application, anomaly, operating system, device, or any other packet grouping.
Attaching a sampleID to every packet in a PCAPNG traffic capture, \sys{}
removes ambiguity surrounding which packets are meant to be associated with a
given traffic sample, eliminating all ambiguity arising from vague traffic
analysis terms that can be interpreted in multiple ways such as ``traffic
sessions'', ``traffic flows'', and ``applications.'' 

\paragraph{Reproducible.}
The sampleIDs that \sys{} generates provide a means for future researchers to
definitively know which traffic samples (\ie, groups of packets) were used for
a given analysis task. This capability greatly enhances reproducibility of
research, as researchers can know that they are able to work on
``apples-to-apples'' methodological comparisons at the packet-level. For
example, sampleIDs enable researchers to publish the sampleIDs of their
training, testing, and validation sets for machine learning based traffic
analysis tasks, allowing future researchers to not only ensure they are using
the same packets in each traffic sample, but even the same dataset splits.
Finally, sampleIDs can increase the depth of understanding across various
methods, allowing researchers to uniquely identify the specific samples on
which their techniques perform well or poorly.

\subsection{Design and Usage} 

We have described the methods that \sys{} uses to satisfy the requirements
outlined in Section~\ref{sec:req}. Figure~\ref{fig:pcapml} shows an overview of
\sys{}, an open source system for standardizing traffic analysis datasets. In
this section outline the capabilities and uses of \sys{} in practice using a
public dataset to provide illustrative examples.

We leverage the Snowflake Fingerprintability dataset~\cite{snowflake} to
walkthrough \sys{}. The dataset contains over 6,500 DTLS handshakes collected
to evaluate the indistinguishability of Snowflake, a pluggable-transport for
Tor that leverages WebRTC, with handshakes from other WebRTC applications:
Facebook messenger, Google Hangouts, and Discord. As with many publicly
released network traffic datasets, the Snowflake dataset was originally
released with raw traffic files separated from metadata: a list of PCAP files,
one PCAP file for each traffic sample (handshake), and a CSV that maps the
traffic in each file to the application that generated the handshake.

\paragraph{Traffic Inputs.}
We design \sys{} to receive and process three types of raw traffic inputs:
\begin{itemize}
        \item A single PCAP.
        \item A directory of PCAPs.
        \item Live traffic.
\end{itemize}
When processed by \sys{}, all input types results in a single, portable output
file: a metadata-encoded PCAPNG file. The Snowflake dataset task corresponds to
labeling a directory of PCAPs. Listing~\ref{lst:snowflake} shows a sample of
the Snowflake dataset on disk.

\begin{lstlisting}[caption={A sample of the DTLS dataset in its originally released format: one PCAP per traffic sample.}, label={lst:snowflake}]
$ ls dataset/
ubuntu_chrome_discord_0.pcap
ubuntu_chrome_discord_1.pcap
...
ubuntu_chrome_facebook_0.pcap
ubuntu_chrome_facebook_1.pcap
...
ubuntu_firefox_snowflake_0.pcap
ubuntu_firefox_snowflake_1.pcap
...
ubuntu_firefox_google_0.pcap
ubuntu_firefox_google_1.pcap
...
\end{lstlisting}

\paragraph{Metadata Inputs.}
\sys{} ingests a metadata file along with the raw traffic inputs in order to
attach metadata to each traffic sample (i.e., group of packets) in a given
dataset. We design \sys{} to accept metadata files in \sys{} follow a
consistent CSV format, with each record containing two or three columns:

\begin{itemize}
    \item \texttt{traffic\_filter}, which designates a filter that a set of
        packets will match and generally represents a single traffic sample.
    \item \texttt{metadata} designates the metadata that will be attached to
        each packet that matches a specific \texttt{traffic\_filter}.
    \item \texttt{group\_key}, an optional third column that enables users to
        generate traffic samples out of multiple \texttt{traffic\_filters} by
        overriding the default sampleID generation and associating all of the
        packets with the same \texttt{group\_key} with a single sampleID.
\end{itemize}

\paragraph{Traffic Filters.}
We have implemented three types of traffic filters in \sys{}:
\begin{itemize}
    \item File: file filters are used when running \sys{} on a directory of
        PCAPs and map all the traffic in a single PCAP file to a piece of
        metadata.
    \item BPF: BPF filters can be used to filter traffic, where every packet
        that matches the BPF filter is associated with a piece of metadata.
    \item Timestamps: timestamp filters map all of the traffic before a given
        timestamp, after a given timestamp, or between two timestamps to a
        piece of metadata.
\end{itemize} 

\sys{} also has the capability to combine BPF and timestamp filters. An example
of the metadata file required to encode the Snowflake dataset is shown in
Listing~\ref{lst:metadata}. Each filter in the \texttt{traffic\_filter} column
is simply prepended by the filter type.

\begin{lstlisting}[caption={An example metadata file when using pcapml on a directory of pcaps.}, label={lst:metadata}]
$ cat metadata.csv
# traffic_filter,metadata,group_key
FILE:dataset/ubuntu_chrome_discord_0.pcap,discord,
FILE:dataset/ubuntu_chrome_discord_1.pcap,discord,
...
FILE:dataset/ubuntu_chrome_facebook_0.pcap,facebook,
FILE:dataset/ubuntu_chrome_facebook_1.pcap,facebook,
...
FILE:dataset/ubuntu_firefox_snowflake_0.pcap,facebook,
FILE:dataset/ubuntu_firefox_snowflake_1.pcap,facebook,
...
FILE:ubuntu_firefox_google_0.pcap,google,
FILE:ubuntu_firefox_google_1.pcap,google,
...
\end{lstlisting}

\paragraph{\sys{} Operation.} When run, \sys{} first parses the metadata file
and generates a vector of traffic filters to be held in memory. For each
traffic filter, \sys{} generates a unique sequential integer sampleID starting
from zero. \sys{} originally used a hashing function to generate sampleIDs for
each traffic sample, but hashing has drawbacks: shorter output lengths leave
the opportunity for hash collisions, while longer output lengths lead to
inflated output file size, as a sampleID is attached to each individual packet.

Next, \sys{} reads in packets from the given input source, searching the filter
vector for matches. \sys{} employs two methods for searching for a filter
match: when matching FILE filters (while processing a directory of PCAPs),
\sys{} loads in the sampleID and metadata associated with all of the packets in
the file before processing in order to avoid vector lookup overhead for each
packet. Conversely, when matching BPF or timestamp filters, \sys{} linearly
searches the filter vector until it finds a match. We implemented a small
optimization in the vector search by beginning the search from the last matched
filter, rather than searching the vector from the beginning for all packets.
This optimization takes advantage of the bursty nature of traffic from hosts or
in flows, allowing for quick lookups for adjacent packets that match identical
filters without adding significant system complexity. 

Finally, \sys{} outputs each packet that matches a traffic filter to the
designated output file, leveraging the PCAPNG packet block options to encode
the sampleID and metadata associated with the packet in a
\texttt{sampleID,metadata} UTF string. Listing~\ref{lst:one_liner} shows that
we can encode the Snowflake Fingerprintability dataset with metadata using a
single command.

\begin{lstlisting}[caption={Encoding a datset with metadata using pcapml.}, label={lst:one_liner}]
$ pcapml -D dataset/ -L metadata.csv -W dataset.pcapng
\end{lstlisting}

\paragraph{\sys{} Output.} \sys{} outputs PCAPNG traffic capture files that
directly couple metadata and raw network traffic while meeting the portability
and compatibility requirements. The result of the above command can still be
parsed by popular tools such as \texttt{tcpdump} and \texttt{tshark} as shown
in Listing~\ref{lst:tcpdump}.

\begin{lstlisting}[caption={pcapml-encoded files are portable to other tools.}, label={lst:tcpdump}]
$ tcpdump -r dataset.pcapng 
23:28:07.118693 IP 74.125.250.71.19305...
23:28:07.119460 IP 192.168.7.222.55937...
23:28:07.142124 IP 74.125.250.71.19305...
23:28:07.143005 IP 192.168.7.222.55937...
22:14:19.944334 IP 74.125.250.26.19305...
22:14:19.945955 IP 192.168.7.222.54537...
22:14:19.971409 IP 74.125.250.26.19305...
22:14:19.972218 IP 192.168.7.222.54537...
23:12:42.033739 IP 74.125.250.71.19305...
23:12:42.036166 IP 192.168.7.222.54510...
...
\end{lstlisting}

Using tools with advanced PCAPNG functionality, such as \texttt{tshark}, we can
easily inspect the metadata encoded into the traffic capture. As shown in
Listing~\ref{lst:tshark}, we see the sampleID for each packet is associated
with (in this case, a DTLS handshake), followed by the label for the traffic
sample. 

\begin{lstlisting}[caption={tools such as tshark can be used to inspect or extract pcapML encoded datasets.}, label={lst:tshark}]
$ tshark -r dataset.pcapng -T fields -e frame.comment
0,google
0,google
0,google
0,google
1,google
1,google
1,google
1,google
2,google
2,google
...
\end{lstlisting}

\paragraph{Sorting by SampleID.} By default, \sys{} outputs packets in the
order they are processed. When processing a directory of PCAPs using file
filters, the packets are naturally sorted by the sampleID as the traffic
samples are already split before being ingested. However, when attaching
metadata to a single PCAP or to live traffic, the output PCAPNG is in
timeseries order by default. As it is often beneficial to sort the packets
first by sampleID, and then by time order; \sys{} provides this functionality,
shown in Listing~\ref{lst:sort}.

\begin{lstlisting}[caption={Sorting a pcapML encoded dataset.}, label={lst:sort}]
$ pcapml -M unsorted_dataset.pcapng -s -W sorted_dataset.pcapng
\end{lstlisting}

\paragraph{Backwards Compatibility.}
Although \sys{}-encoded PCAPNGs are portable across many different tools and
tasks, not every traffic analysis system supports the PCAPNG traffic capture
format. As such, we have implemented functionality in \sys{} to revert
\sys{}-encoded PCAPNGs out to PCAPs, using one PCAP file per traffic sample. 

When reverting, \sys{} encodes the sampleID and metadata for each traffic
sample in the name of each PCAP file and outputs a metadata CSV file containing
a full mapping of PCAP files to their associated metadata.

\subsection{\sys{} Overhead and Performance} In this section we evaluate the
performance of \sys{} in a variety of dimensions, including the time it takes
to label a variety of datasets, the cost (in disk space) of attaching
information to every packet in a traffic capture, and the speed at which the
public implementation of \sys{} can label live traffic.

\paragraph{\sys{} Metadata Overhead.}

\input{figures/tables/pcapml/pcap.tex}
\input{figures/tables/pcapml/dir.tex}

We first examine the overhead file size cost of \sys{} by setting up an
experiment where we increase the size of the metadata attached to each packet
in a given traffic capture. For this experiment, we leverage \texttt{tcpreplay}
to replay traffic and use the \texttt{bigFlows.pcap} capture to provide a
traffic example~\cite{tcpreplay,tcpreplay-flows}. The traffic capture contains
almost 800,000 packets across over 40,000 traffic flows and 132 applications
with an average packet size of 449 bytes.

We select 20 random source IP addresses (out of a possible 3,218 IP addresses
in the capture) from the capture to tag with metadata. We use the BPF filtering
mechanism in \sys{} to tag all of the traffic originating from the selected IP
addresses, attaching a static number of metadata bytes to each packet matching
our filters. Finally, we strip the resulting \sys{}-encoded PCAPNG of the
metadata, transforming it into a PCAP in order to calculate the disk overhead
due to the metadata. We run the experiment 10 times for each metadata size
tested. 

Table~\ref{tab:pcap-eval} shows the results of this experiment. As expected, we
see that the disk overhead of \sys{} increases with the size of the metadata to
be attached to each packet. We also see that the minimum and maximum size
increases across the traffic varies significantly.

Ultimately, a number of factors influence the size of the \sys{}-encoded PCAPNG
file when compared to a PCAP without metadata. First, the overhead of \sys{}
can increase due to a large number of samples in a dataset. The number of bytes
required to write higher sampleID values to file increases with the number of
samples to be encoded. Second, the distribution of packet sizes can influence
the overhead cost of \sys{}. A small number of large packets will incur
relatively lower overhead than a large number of smaller sized packets.
Finally, the size of the metadata to be attached to the packet will influence
the overhead cost of \sys{}. As the metadata is attached to each packet with
the associated sampleID, larger metadata can increase disk space overhead.

\paragraph{\sys{} Overhead On Public Datasets.} Next, we evaluate \sys{} on
eight public datasets. These eight datasets were published by
Holland~\etal{}~\cite{nprint} and represent both 1) a variety of tasks:
including device identification, application identification, intrusion
detection, and OS detection; and 2) a variety of traffic types: containing
various types of packets from DTLS traffic handshakes to UDP and TCP traffic.

Table~\ref{tab:pcapml-dir-eval} examines the overheads and performance of using
\sys{} on these eight datasets. First, we see that each dataset was originally
released as a set of PCAPs, one PCAP per traffic sample.
Table~\ref{tab:pcapml-dir-eval} includes both the number of files in the
originally formatted dataset and the size of the dataset on disk as reported by
\texttt{du}. Next, we see the size of the \sys{}-encoded PCAPNG file after
processing the files using \sys{}. In many cases, the size of the dataset
decreases significantly---this is a product of reducing the number of files on
disk, as filesystem block size overheads dominate datasets with large numbers
of files. In other cases, the size of the dataset increases on disk when
attaching metadata using \sys{}. 

Table~\ref{tab:pcapml-dir-eval} demonstrates the overhead cost across the eight
encoded datasets by stripping the \sys{}-encoded PCAPNG of the attached
metadata and transforming it into a PCAP traffic capture. The cost of attaching
the sampleID and metadata to each of these datasets ranges from 5 to 30\%,
depending on the factors previously discussed. Finally,
Table~\ref{tab:pcapml-dir-eval} reports the amount of time \sys{} took to
encode the dataset and decode the dataset for backwards compatibility. We see
that even the largest dataset, containing 18GB of traffic, takes less than a
minute to encode and decode using \sys{}.

\begin{figure}[t]
    \centering
    \includegraphics[scale=0.50]{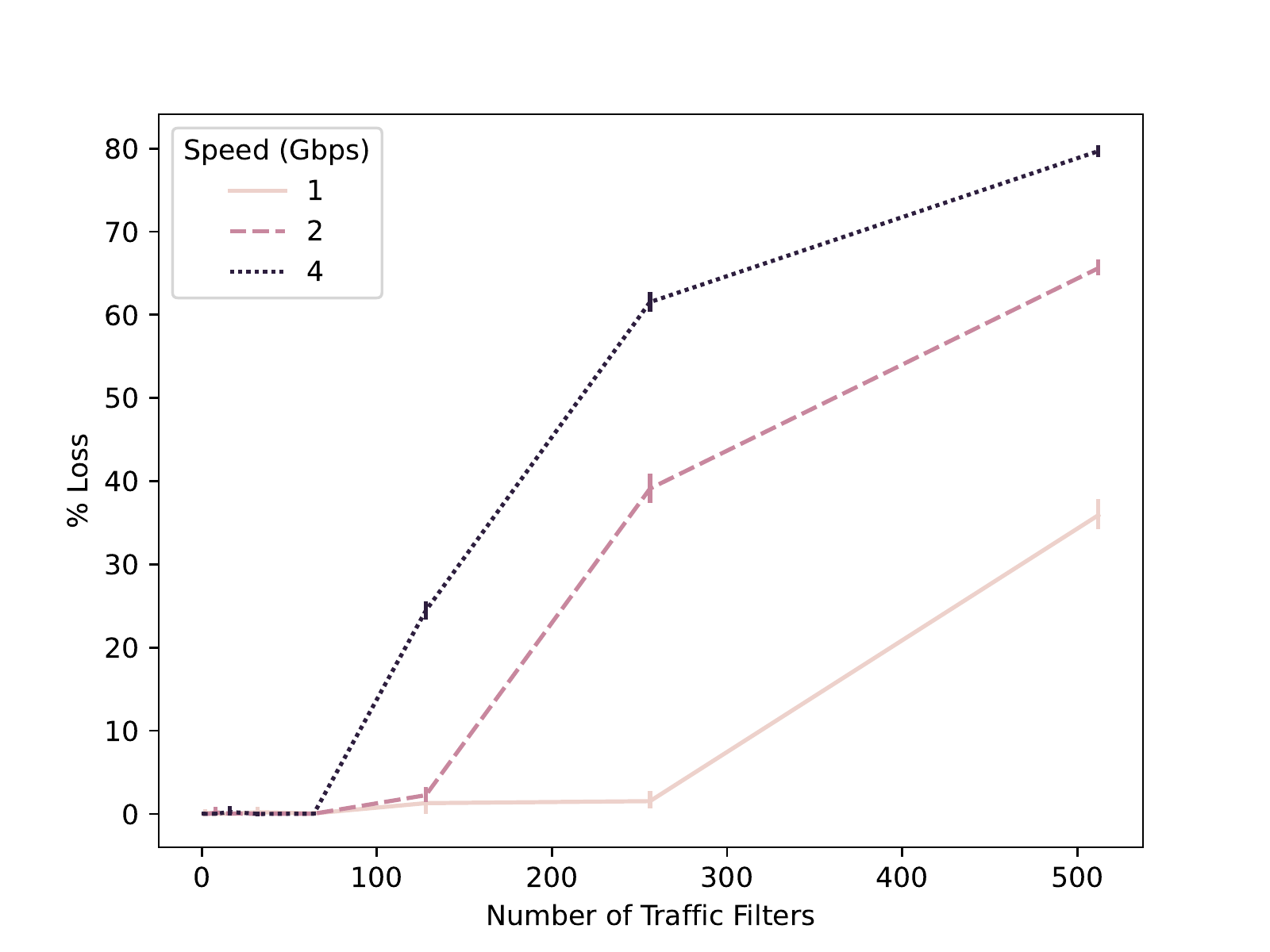}
    \caption{\sys{} can encode metadata into live traffic in configurations that 
    satisfy many experimental use cases.}
    \label{fig:speed}
\end{figure}


\paragraph{\sys{} Performance on Live Traffic.} 

\sys{} can capture and attach
metadata to traffic on a live network interface. This functionality can be
useful for researchers who wish to run traffic experiments as they can capture,
label, and ID the traffic samples for their experiment in real time. We measure
the speeds which \sys{} can capture and attach metadata to incoming packets
while avoiding packet losses. \sys{} leverages \texttt{libpcap} to read packets
from an interface and uses a custom PCAPNG writer for matching and writing
packets to disk. The existing implementation of \sys{} uses a single
thread.\footnote{We anticipate that \sys{} could be tuned and refactored for
high performance network environments using optimized packet capture libraries
and concurrency. We leave this for future work.} 

We again leverage \texttt{tcpreplay}, and the same 
\texttt{bigFlows.pcap} traffic capture as our previous experiment examining
disk overhead. We vary two input parameters. First, we vary the number of
samples that \sys{} is attempting to tag. Recall that, for each packet, \sys{}
must check each traffic filter in the vector until it finds a match. As such,
the time \sys{} takes to check for a match increases with the number of filters
in the search vector. Next, we vary the speed at which packets are replayed on
the interface to understand the processing capabilities of \sys{} and at which
rates losses occur.

We run each individual (number samples, speed) configuration five times and
report the recorded loss of each combination. Figure~\ref{fig:speed} examines
the results of the experiment in detail, showing the loss for each
configuration tested. We see that \sys{} can generally filter and tag live
traffic at 1 Gbps with no loss for up to 256 defined traffic samples. \sys{}
can tag up to 100 samples with zero loss, but struggles with larger numbers of
samples. Finally, \sys{} can tag live traffic at 4 Gbps for a lower number of
samples (up to 64) with roughly zero loss. We note here that for many use
cases, such as capturing and labeling website traffic or device traffic, the
number of filters used in practice is likely much lower than 64. For example,
capturing and labeling traffic for a 5-tuple can be done in a single traffic
sample using the BPF filter functionality.

\subsection{Enabling Analysis in Existing Pipelines with \fe{}} 

The ultimate goal of many traffic analysis tasks is to extract information from
a set of packets to identify sets of traffic samples, such as identifying
different devices on a network. A non-negligible portion of traffic analysis
work is done using \texttt{python}, along with several popular libraries. To
facilitate exploration of new analysis methods and incorporation of \sys{} into
existing pipelines, we have developed \fe{}, an open source tool which
leverages the standardized dataset format of \sys{} to enable researchers to
focus on analysis, rather than dataset formatting.

\fe{} is a python modules that interfaces directly with \sys{} encoded
datasets, exposing an iterator over individual traffic samples and their
associated metadata. \fe{} can transform packets into popular python packet
analysis libraries such as \texttt{scapy} and \texttt{dpkt}~\cite{scapy,dpkt}
in a single line of code. Finally, we point out that one of the most important
capabilities that \fe{} and \sys{} provides is simplifying the search for
generalizable traffic analysis methods. The standardized dataset format of
\sys{} and standard interface of \fe{} enables researchers to test methods
across datasets and tasks by simply loading in a new dataset. Below is an
example of \fe{} loading and iterating over the Snowflake Fingerprintability
dataset encoded with \sys{}.

\begin{lstlisting}[caption={\fe{} enables researchers to focus on analysis methods. Testing
an analysis method on a different dataset only requires loading in a different pcapML encoded dataset.}, label={lst:fe}]
import argparse
import pcapml_fe
from pcapml_fe_helpers import *

def main():
    parser = argparse.ArgumentParser()
    parser.add_argument('pcapml_dataset')
    args = parser.parse_args()
    
    for traffic_sample in pcapml_fe.sampler(args.pcapml_dataset):
        analysis_method(traffic_sample)

def analysis_method(sample):
    print(sample.sid)
    print(sample.metadata)
    
	for pkt in traffic_sample.packets:
        # Raw bytes and timestamp
        print(pkt.ts, pkt.raw_bytes)
        # Transform to Scapy packet in one line 
        spacket = scapy_readEther(pkt.raw_bytes)
\end{lstlisting}

%% file: figures/tables/pcapml/pcap.tex
\begin{table}[t]
    \centering
    \small
\begin{tabular}{@{}rrrr@{}}
\toprule
\multicolumn{1}{l}{}                                                                & \multicolumn{3}{c}{\% Increase over PCAP}                                    \\ \cmidrule(l){2-4} 
\multicolumn{1}{c}{\begin{tabular}[c]{@{}c@{}}Metadata Size\\ (Bytes)\end{tabular}} & \multicolumn{1}{c}{Mean} & \multicolumn{1}{l}{Min} & \multicolumn{1}{c}{Max} \\ \midrule
\rowcolor[HTML]{C0C0C0} 
1                                                                                   & 8.7                      & 5.1                     & 12.3                    \\
5                                                                                   & 8.1                      & 3.8                     & 13.3                    \\
\rowcolor[HTML]{C0C0C0} 
10                                                                                  & 10.3                     & 4.6                     & 19.1                    \\
20                                                                                  & 12.0                     & 4.6                     & 24.2                    \\
\rowcolor[HTML]{C0C0C0} 
40                                                                                  & 16.3                     & 6.5                     & 29.4                    \\
80                                                                                  & 19.3                     & 11.1                    & 27.8                    \\
\rowcolor[HTML]{C0C0C0} 
160                                                                                 & 30.3                     & 17.1                    & 51.1                    \\ \bottomrule
\end{tabular}
\caption{The disk overhead of pcapML encoded traffic captures generally increase with the size of the metadata.}
\label{tab:pcap-eval}
\end{table}

%% file: figures/tables/pcapml/dir.tex
\begin{table*}[t]
    \centering
    \resizebox{\textwidth}{!}{
\begin{tabular}{lrrrrrrr}
\hline
\multicolumn{1}{c}{Dataset} & \multicolumn{1}{c}{\# Pcap Files} & \multicolumn{1}{c}{Disk Size} & \multicolumn{1}{c}{\begin{tabular}[c]{@{}c@{}}PcapML-Encoded
\\ Size\end{tabular}} & \multicolumn{1}{c}{Stripped Pcap Size} & \multicolumn{1}{c}{\begin{tabular}[c]{@{}c@{}}\% Increase Over \\ Raw PCAP\end{tabular}} & \multicolumn{1}{c}{\begin{tabular}[c]{@{}c@{}}Time To Encode\\ (Seconds)\end{tabular}} & \multicolumn{1}{c}{\begin{tabular}[c]{@{}c@{}}Time To Split\\ (Seconds)\end{tabular}} \\ \hline
\rowcolor[HTML]{C0C0C0} 
Active Case Study           & 274,009                           & 1.1 GB                            & 30 MB                                                                              & 25 MB                         & 17                                                                                       & 3                                                                                      & 7                                                                                     \\
Application Case Study      & 5,787                             & 33 MB                             & 20 MB                                                                              & 18 MB                         & 10                                                                                       & 0.2                                                                                    & 0.4                                                                                   \\
\rowcolor[HTML]{C0C0C0} 
Cross Market Case Study     & 521                               & 8.6 GB                            & 8.8 GB                                                                             & 8.4 GB                        & 5                                                                                        & 12                                                                                     & 21                                                                                    \\
netML IDS                   & 558,884                           & 12 GB                             & 8.0 GB                                                                             & 7.4 GB                        & 8                                                                                        & 17                                                                                     & 32                                                                                    \\
\rowcolor[HTML]{C0C0C0} 
netML IoT                   & 498,446                           & 6.7 GB                            & 5.9 GB                                                                             & 5.3 GB                        & 10                                                                                       & 13                                                                                     & 30                                                                                    \\
netML Type of Traffic       & 158,355                           & 18 GB                             & 13 GB                                                                              & 12 GB                         & 8                                                                                        & 39                                                                                     & 46                                                                                    \\
\rowcolor[HTML]{C0C0C0} 
OS Case Study               & 124,390                           & 529 MB                            & 241 MB                                                                             & 187 MB                        & 22                                                                                       & 2                                                                                      & 3                                                                                     \\
Video Case Study            & 20,980                            & 330 MB                            & 414 MB                                                                             & 288 MB                        & 30                                                                                       & 2                                                                                      & 5                                                                                     \\ \hline
\end{tabular}
    }%
\caption{pcapML can encode datasets in an efficient amount of time with overhead cost depending on the amount of metadata to be attached.}
\label{tab:pcapml-dir-eval}
\end{table*}

%% file: sections/benchmarks.tex
\section{\sys{} Benchmarks}\label{sec:benchmark}

\sys{} represents a ``narrow waist'' in network traffic analysis, providing
a standardization solution for storing and processing network traffic 
datasets. Such a narrow waist can be used for broader impact by leveraging 
standardization to centralize and track progress in the field. Along with 
\sys{}, we have created the \sys{} benchmarks~\footnote{nprint.github.io/benchmarks}, 
an open source repository and public leaderboard with the goal of centralizing 
and standardizing network traffic analysis research. 
As of writing, the \sys{} benchmarks 
contain seven datasets across five discrete tasks, including website 
fingerprinting, device identification, malware detection, and application
identification.

\textit{Any researcher or practitioner} can add a new task, a new dataset, 
or submit results of a method on one of the public leaderboards. The only 
requirement being that each submitted task and dataset
be encoded with \sys{} for standardization. For dataset
curators, the \sys{} benchmarks provide a central repository to list their
datasets, tasks, and any initial results on the dataset, without having to
develop custom methods for storing and releasing each dataset. For 
dataset users, the \sys{} benchmarks ensure any methods developed
and evaluated can be directly compared with any other technique using the 
same \sys{} benchmark. Further, a central repository of 
standardized datasets enables dataset users to easily test methods on a
variety of traffic analysis tasks, encouraging researchers to create and 
evaluate \textit{generalizable} methods. Finally, for the field, a centralized
benchmark repository provides an avenue for tracking the progress 
of techniques, making it easier to discern when a methodological breakthrough
has occurred.

%% file: sections/whereto.tex
\section{General Recommendations}\label{sec:gen}

We have highlighted issues that simultaneously inhibit reproducibility and
innovation in network traffic analysis. While \sys{} is an initial solution for
reproducible network traffic analysis, it is not a panacea. Moving forward, we
urge the community to consider approaching this research area with a broader
lens. Here we discuss the different community stakeholders and their role in
making network traffic analysis research more reproducible, and subsequently
more impactful.

\paragraph{Dataset Creators.}
Dataset creators curate and release network traffic datasets for the community
to use, inherently guiding downstream research. The community must converge on
a standard dataset release format to enable sound research and lower the
barrier for new techniques to be developed. We have introduced \sys{} as a
possible avenue for generating standard network traffic datasets, but any
format which meets the requirements outlined in Section~\ref{sec:req} would
improve upon the current status quo.

\paragraph{Dataset Users.}
Dataset users play an equally important role in improving reproducibility in
network traffic analysis. Many traffic analysis techniques report high levels
of performance on a given task (\eg, $\geq 90$ precision) such as website
fingerprinting or intrusion detection. As a field, we are entering an era of
diminishing returns as we near the point in which many existing methods and
techniques perform ``well enough'' for most tasks. New methods 
are likely to represent relatively small, yet vitally important performance
increases. For instance, even performance increases of 1\% can denote a
methodological advancement. However, without \textit{exact} comparisons of
methods, such small performance increases can be difficult to ascribe to a
method versus other factors. A set of best practices when
performing analysis will enable the field to better understand when a
methodological advancement has occurred.

First, we must focus efforts on testing methods on full datasets when possible.
Due to the large number of samples found in many network traffic datasets, it
has become accepted to evaluate methods on sampled versions of full network
traffic datasets. This practice can be dangerous due to the temporal nature of
network traffic (\ie, events are not necessarily independent). If a dataset is
sampled to evaluate a method, either the sampled dataset must be re-encoded and
released, or in the case of \sys{}, the sampleIDs used in evaluation can easily
be released. If machine learning techniques are used for a task, releasing
specific training and testing datasets (or sampleIDs) will help to directly 
compare methods. Third, if a dataset is leveraged for a purpose other than its
original release, such as re-labeling the traffic for a different task, the
dataset should be re-encoded and released to guarantee reproducible research.
Finally, when releasing pre-processed data, such as features used to train a
model, a direct link to the raw network traffic should be traceable. For
example, releasing a CSV of features which includes a sampleID column will
enable future work to better contextualize new methods with old ones.

%% file: sections/related.tex
\section{Related Work}\label{sec:related}

This section explores recent standardization and reproducibility work as well
as work related to \sys{}.

\paragraph{PCAPNG.}
The PCAPNG traffic capture format has been in development since 2014. As such,
research has leveraged the format in a variety of ways. Le~\etal{} leverage the
PCAPNG file format in AntMonitor, a system for monitoring mobile devices, to
attach mobile application names to raw packets~\cite{le2015antmonitor}.
Velea~\etal{} also leverage the PCAPNG traffic capture format to encode
pre-processed feature information, such as the use of encryption, the protocol,
and the number of packets in a flow using a custom-developed block
option~\cite{velea2017network,velea2017feature}. In contrast to these works,
our work focuses on building a \textit{generalizable} system for network
traffic analysis tasks by encoding arbitrary metadata onto packets.

\paragraph{Reproducibility and ML}
A trove of recent research has examined and highlighted irreproducibility 
across a variety of fields. Most famously, the Open Science Collaboration
highlighted the reproducibility crisis in psychological
science~\cite{open2015estimating}. More recently, much work has focused on the
reproducibility of applied machine learning research. Applied machine learning
reproducibility failures have been brought to light in fields including:
neuroimaging, bioinformatics, medicine, software engineering, toxicology,
radiology, epidemiology, political science, and
nutrition~\cite{whelan2014optimism,kapoor_irreproducible_2021,bone2015applying,
lusa2015joint,ivanescu2016importance,olorisade2017reproducibility,ye2018prediction,
tu2018careful,christodoulou2019systematic,dacrema2021troubling,alves2019oy,mcdermott2021reproducibility,
vandewiele2021overly,roberts2021common}. 

Most related to our work is Arp~\etal{}'s examination of applied machine
learning in a variety of computer security research\cite{arp2020and}.
Arp~\etal{} examine the use of incorrect application of machine learning
techniques to a single intrusion detection dataset, demonstrating the methods
used in Mirsky~\etal{} are likely overly complex for the
task~\cite{mirsky2018kitsune}. In contrast, our work examines the area of
network traffic analysis more generally, demonstrating that any analysis
performed, not only machine learning, is irreproducible due to issues existing
\textit{before} the analysis stage of a pipeline.

%% file: sections/conclusion.tex
\section{Conclusion}\label{sec:conclusion}

In this work we highlighted a reproducibility crisis facing network traffic
analysis research. We examined the usage of multiple popular datasets,
highlighting how the existing ecosystem leads to researchers testing methods on
different versions of the same dataset. We then inspected the literature to
outline the barriers to reproducibility in the field, which we used to develop
a list of standardization requirements.

Given our findings, we introduced \sys{}, an open source system which meets the
outlined requirements. \sys{} standardizes dataset curation and usage by
enabling researchers to directly encode metadata into raw traffic traces,
eliminating ambiguous language and engineering errors that stem from the
variety of dataset formats that currently exist. We evaluate \sys{} across
multiple dimensions, including the increase in dataset size when encoding
traffic datasets using \sys{} and \sys{}'s capability to encode metadata into
traffic in real-time. Finally, we demonstrate the broader impact that 
standardization can have on the field by creating the \sys{} benchmarks, 
a public leaderboard website and repository built to track the progress
of methods in network traffic analysis. Ultimately, we see \sys{} and the 
\sys{} benchmarks as an avenue to propel the field forward, enabling rapid
innovation through centralization and standardization.